\documentclass[aps,prb,twocolumn,superscriptaddress,showpacs]{revtex4}
\pdfoutput=1
\usepackage{amssymb}
\usepackage{upgreek}
\usepackage{graphicx}
\usepackage{mathrsfs}
\usepackage[utf8]{inputenc}
\usepackage{color}

\begin{document}

\title{Metastability and hysteretic vortex pinning near the order-disorder transition in NbSe$_2$: An interplay between plastic and elastic energy barriers?}

\author{M. Marziali Bermúdez}
\affiliation{Departamento de Física, Facultad de Ciencias Exactas y Naturales, Universidad de Buenos Aires, Argentina.}
\affiliation{Instituto de Física de Buenos Aires, Consejo Nacional de Investigaciones Científicas y Técnicas, Argentina.}

\author{E. R. Louden}
\affiliation{Deptartment of Physics, University of Notre Dame, Notre Dame, IN 46556, USA.}

\author{M. R. Eskildsen}
\affiliation{Deptartment of Physics, University of Notre Dame, Notre Dame, IN 46556, USA.}

\author{C. D. Dewhurst}
\affiliation{Institut Laue-Langevin Avenue des Martyrs BP156, 38042 Grenoble Cedex 9, France.}

\author{V. Bekeris}
\affiliation{Departamento de Física, Facultad de Ciencias Exactas y Naturales, Universidad de Buenos Aires, Argentina.}
\affiliation{Instituto de Física de Buenos Aires, Consejo Nacional de Investigaciones Científicas y Técnicas, Argentina.}

\author{G. Pasquini}
\email{pasquini@df.uba.ar}
\affiliation{Departamento de Física, Facultad de Ciencias Exactas y Naturales, Universidad de Buenos Aires, Argentina.}
\affiliation{Instituto de Física de Buenos Aires, Consejo Nacional de Investigaciones Científicas y Técnicas, Argentina.}

\date{\today }


\begin{abstract}
We studied thermal and dynamic history effects in the vortex lattice (VL) near the order-disorder transition in clean $\mathrm{NbSe}_2$ single crystals. Comparing the evolution of the effective vortex pinning and the bulk VL structure, we observed metastable superheated and supercooled VL configurations that coexist with a hysteretic effective pinning response due to thermal cycling of the system. A novel scenario, governed by the interplay between (lower) elastic and (higher) plastic energy barriers, is proposed as an explanation for our observations: Plastic barriers, which prevent the annihilation or creation of topological defects, require dynamic assistance to be overcome and to achieve a stable VL at each temperature. Conversely, thermal hysteresis in the pining response is ascribed to low energy barriers, which inhibit rearrangement within a single VL correlation volume and are easily overcome as the relative strength of competing interactions changes with temperature.
\end{abstract}

\maketitle

\section{Introduction} 
\label{sec:intro}

A wide variety of complex systems can be described as collectively interacting particles on random potential landscapes. Among others, vortices in type-II superconductors,\cite{abrikosov} charge density waves,\cite{cdw} electron crystals,\cite{elxtal} colloids,\cite{colxtal} and skyrmions\cite{skyrm} belong to this category. History effects related to plasticity, glassy behavior and metastable configurations in the vicinity of an order-disorder transition (ODT) are common features in these systems.\cite{Reviews} In this context, vortex matter is a prototype model where vortex-vortex interactions, favoring an ordered vortex lattice (VL), compete with both thermal fluctuations and pinning interactions that tend to disorder the system, resulting in multiple vortex phases and complex dynamics. In the last decades, a large amount of work has been dedicated to the study of vortex matter in the presence of strong pinning, both in conventional and non-conventional superconducting materials. Such systems are promising for practical transport applications.\cite{transport} On the other hand, from a fundamental point of view,\cite{Blatter1994} interesting physics arises when the inter-vortex interaction plays a major role, competing with a weak vortex-pinning interaction.\cite{Blatter1994,Giammarchi1995}  

In a variety of superconducting materials, characterized by the presence of randomly distributed weak pinning centers, the stable vortex phase at low temperature and low magnetic fields is an ordered Bragg Glass (BG) phase without dislocations.\cite{Giammarchi1995, Klein2001} Yet, when the system is cooled from the normal state in an external magnetic field (field-cooled [FC]), energy barriers trap the VL in disordered metastable configurations where the VL is more strongly pinned.\cite{Gammel1998} By applying high transport current densities \cite{Gammel1998, Henderson1996, Paltiel2000} or large oscillatory shaking magnetic fields,\cite{Pasquini2008, Daroca2011} the system overcomes energy barriers and reaches the ordered BG phase, which has lower effective pinning. This ordered phase is characterized by the weak elastic decay of spatial correlations over large distances, coexisting with a glassy behavior. With increasing field and/or temperature, the system undergoes an ODT to a disordered glass\cite{Giammarchi1997, Gammel1998} with correlation lengths $\zeta$ of the order of the mean inter-vortex distance, $a_{0}$. The sudden increase in the effective pinning at the ODT generates the well-known Peak Effect (PE) anomaly.

In very low pinning materials, such as clean $\mathrm{NbSe_{2}}$ single crystals, the BG phase is stable in a large region of the vortex field-temperature phase diagram.\cite{Du2007} Here, the ODT takes place near the superconducting phase boundary.\cite{Xiao2000} Intermediate responses observed in transport experiments in the region of the PE, indicating an intermediate degree of pinning, had been originally ascribed to a contamination produced by the incoming disordered vortices trough the sample edges.\cite{Paltiel2000} However, the existence of a narrow transitional region between the BG and the disordered phase, in the vicinity of the PE, has been theoretically proposed \cite{Menon2012} and experimentally confirmed. Recently, by joint small-angle neutron scattering (SANS) and in-situ AC susceptibility measurements, we were able to show that the application of a shaking AC field in the transitional region drives the VL to bulk robust configurations with intermediate degrees of disorder.\cite{Marziali2015} Numerical simulations \cite{Daroca2011} suggested that these intermediate configurations could arise from stationary fluctuating dynamic states, similar to those observed in colloidal systems.\cite{Pine2008, Okuma2011} Stable VL configurations with intermediate dislocation density have also been observed in recent scanning tunneling spectroscopy (STS) experiments carried out in the transitional region of irradiated \cite{Zehet2015} and Co-doped \cite{Ganguli2015} $\mathrm{NbSe_{2}}$ single crystals.

In our previous experiment, in-situ linear AC measurements revealed that configurations with intermediate degrees of disorder, created through different dynamic histories, display an intermediate effective pinning.\cite{Marziali2015} The observed correlation between stronger effective pinning and shorter spatial correlation lengths (higher dislocation densities) suggested that dynamic history effects in the linear AC response are a consequence of changes in the bulk VL dislocation density. On the other hand, hysteretic effective pinning observed after cooling and/or warming the system without any dynamic assistance are poorly understood up to now. \cite{Pasquini2008, Li2006, Daniilidis2007, Ganguli2016}  Here, we report a systematic study of the effective pinning and structural characterization of the VL after various thermal histories in $\mathrm{NbSe_{2}}$ single crystals. Our results support a novel scenario to describe thermal history effects in vortex matter and the controversial relationship between the VL structure and the effective pinning.

This paper is organized as follows: In section \ref{sec:techniques}, an introduction to the experimental techniques and its relationship with the underlying VL properties is presented. In section \ref{sec:details}, some relevant technical details regarding experiments and samples are given. In section \ref{sec:results}, results corresponding to the AC susceptibility VL pinning measurements and SANS structural observations are presented and discussed. Section \ref{sec:conclusions} summarizes the conclusions.

\section{Experimental techniques and underlying physics}
\label{sec:techniques}

\subsection{Linear AC susceptibility and the Campbell regime}

In order to investigate the spontaneous modification of VL pinning properties, a non-invasive experimental technique is imperative. An ideal way to measure the VL response in weak pinning type-II superconductors is by AC susceptibility restricted to the linear Campbell regime:\cite{Campbell} The application of a very small AC field $h_{\mathrm{AC}}$, superimposed to the DC field $H$ forces vortices to perform small (harmonic) oscillations inside their effective pinning potential wells without modifying their spatial configuration. These oscillations propagate through the sample due to the vortex-vortex repulsion, with a characteristic penetration depth $\lambda _{\mathrm{AC}}=\lambda _{\mathrm{R}}+i\lambda _{\mathrm{I}}$. In the Campbell regime $\lambda_{\mathrm{I}}\ll \lambda _{\mathrm{R}}$ and 
\begin{equation}
\lambda _{\mathrm{R}}^{2}=\lambda _{\mathrm{L}}^{2}+\frac{\phi_{0} B}{4 \pi \alpha_{\mathrm{L}}},
\end{equation}
where $\lambda_{\mathrm{L}}$ is the London penetration depth, $\phi_{0} = h/2e \simeq 2068~\mathrm{T~nm^2}$ is the flux quantum and $\alpha_{\mathrm{L}}$ is the curvature of the effective mean pinning potential, also known as the Labusch constant. Since the real component of the AC susceptibility $\chi'$ is determined by the experimental geometry and the dimensionless parameter $\lambda_{\mathrm{R}} / D$, with $D$ a characteristic sample dimension,\cite{Brandt94} $\chi'$ can be directly related to the effective pinning potential in a mean field approximation.\cite{Colo2014} Lower values of $\chi'$ correspond to stronger pinning, saturating to $\chi' = -1/4\pi $ at perfect AC shielding ($\lambda_{\mathrm{R}}/D\ll 1$).

The highest $h_{\mathrm{AC}}$ allowed in the Campbell regime is restricted by the condition that the mean vortex displacement $u$ induced by the AC field be smaller than the characteristic pinning potential size, estimated to be of the order of the superconducting coherence length $\xi$. The displacement is usually largest near the edge of a sample. When $h_{\mathrm{AC}}||H$, the maximum vortex displacement in a sample of size $D$ (perpendicular to the magnetic field) is roughly $u \sim D~h_{\mathrm{AC}}/H$. Therefore, fields $h_{\mathrm{AC}} < H ~ \xi /D$ guarantee $u < \xi$ throughout the sample, a necessary condition to perform linear non-invasive AC susceptibility measurements. The validity of the Campbell regime can be confirmed by testing the linearity of the frequency independent AC response and the smallness of the dissipative component $\chi''$.\cite{Pasquini99}

\subsection{SANS in vortex matter}
\label{sec:sansvortex}

SANS is a non invasive technique suitable for probing magnetic structures at nanometric scales. Here, the incident neutrons' magnetic moment causes them to be diffracted by the magnetic field modulation due to the VL. Diffracted neutrons are collected using a position-sensitive detector, thus obtaining the scattered intensity distribution on the detector plane.

For a perfectly regular hexagonal lattice with lattice parameter $a_{0}$, reflections will take place only for scattering vectors $\mathbf{Q_0}$ matching any of the reciprocal lattice vectors; thus first order Bragg peaks (BPs) will correspond to $Q_0=\frac{4\pi }{a_{0}\sqrt{3}}$. The scattering process is elastic and the scattering angle $\theta_0$, measured relative to the field direction, is given by Bragg's law:
\begin{equation}
\sin \theta _{0}=\frac{Q_0}{2K},
\end{equation}
where $K = 2\pi/\lambda_\mathrm{N}$ is the  neutron wave vector corresponding to a wavelength $\lambda_\mathrm{N}$. Since $Q_0 \ll  K$, the scattering angle is small and $\sin \theta_0 \approx \theta_0$.

For an imperfect lattice with dislocations, spatial correlations decay exponentially beyond a characteristic length $\zeta$ associated with the spatial scale at which dislocations appear.\cite{Giammarchi1995} The anisotropy of the elastic constants is reflected in different $\zeta_{i}$ for $i=x,y,z$. Due to finite correlation lengths, the Bragg reflections are broadened in reciprocal space, with $\Delta {Q}_{i} \propto \zeta_{i}^{-1}$.\cite{Klein2001} Consequently, Bragg peaks will appear broader in the detector plane as well as for angles around the Bragg condition, $\theta = \theta_{0}$. In addition, the intensity distribution is further smeared by the instrumental resolution. Due to the poorer in-plane resolution, only the longitudinal characteristic length in the direction of the flux lines $\zeta_{z}$ can be reliably determined from the measured intensity distribution $I(\theta)$, as long as $\zeta_{z}$ does not exceed resolution limits. This characteristic length is directly related to the mean distance between edge dislocations.\cite{edgedislocations}

The total scattered intensity depends on the density of scatterers and their individual strength, which is proportional to the square of the individual vortex form factor, $F(Q)$.\cite{ReviewMorten} An approximate expression for $F(Q)$ can be obtained from the solution to the London equation by introducing an \textit{ad-hoc} cut-off due to the finite vortex core size. This leads to\cite{Brandt1997}
\begin{equation}
F(Q) \propto \frac{\exp (-c ~ Q^2 \xi^2 )}{1 + Q^2 \lambda_\mathrm{L}^2},
\label{London model}
\end{equation}
where $\xi$ is the superconducting coherence length and $c$ is a heuristic factor. The modified London model has been successfully applied to fit experimental data corresponding to d-wave or multi-gap superconductors using $c$ between $1/4$ and $2$.\cite{Brandt1997,LondonModel} 

Since $\lambda_\mathrm{L}$ and $\xi$ increase with temperature, the magnetic field modulation smears out and thus $F(Q) \rightarrow 0$ as $T \rightarrow T_\mathrm{c}$. It is therefore not feasible to measure full rocking curves to obtain the intensity distribution $I(\theta)$ at temperatures near the ODT. However, when spatial correlation is broken by dislocations, the integrated intensity is conserved.\cite{Klein2001, ReviewMorten} Therefore, the product between the width $\Delta \theta$ and the maximum intensity $I_\mathrm{max} = I(\theta_{0})$ of a BP remains proportional to $F^{2}$. As a result, differences in $\zeta_{z}$ between configurations at equal $T$ and $H$ can be assessed by comparing their corresponding $I_\mathrm{max}$. Moreover, using independent data for $\lambda_\mathrm{L}(T)$ and $\xi(T)$, it is possible to calculate the expected temperature dependence of $F(Q)$, and thereby trace the evolution of $\zeta_{z}$ with temperature by the deviation of $I_\mathrm{max}(T)$ from the expected $F^{2}$-like dependence.

\section{Experimental details}
\label{sec:details}

\subsection{Samples}

The study has been performed in clean $\mathrm{NbSe_{2}}$ single crystals with $T_{c}\sim 7.2~\mathrm{K}$, grown in Bell Labs as described elsewhere.\cite{BellLabs} The phase diagram for a number of crystals was characterized using a 7-T MPMS XL (Quantum Design) and reported previously.\cite{SIMarziali2015} Our crystals only showed minor, insignificant variations compared to other literature results. Most of the linear AC susceptibility measurements were done using a single $(1 \times 1 \times 0.2)~\mathrm{mm^{3}}$ sample, whereas SANS observations and test in-situ AC susceptibility measurements were performed using a larger $(5.6\times 4.7\times 0.2)~\mathrm{mm^{3}}$ crystal.

\subsection{Experimental setups and technical details}

Linear AC susceptibility experiments were performed using a home-made susceptometer based on the mutual inductance technique where both the AC and DC fields are parallel to the $\mathbf{c}$ axis of the sample. The susceptometer was installed in a bath He cryostat that allowed temperature regulation within $\Delta T \leqslant 2 ~ \mathrm{mK}$ down to $T_{\min} \sim 4.3~ \mathrm{K}$.

 In order to study thermal history effects throughout a wide range of temperatures, experiments were performed at a moderate DC field, $H=1~\mathrm{kOe}$. For all susceptibility measurements, AC field amplitudes were restricted to $h_\mathrm{AC}<10~\mathrm{mOe}$. Linear response and low dissipation, which are characteristic features of the linear Campbell regime, were verified. For some sequences, susceptibility measurements were interrupted at certain target temperatures to apply a shaking AC field. Here, $1000$ cycles of a larger sinusoidal field with amplitude $\geqslant 5~\mathrm{Oe}$ and frequency $1~\mathrm{kHz}$ was applied before resuming linear susceptibility measurements.

The SANS experiments were conducted using the D33 instrument at the Institut Laue-Langevin (ILL).\cite{ILLdata} Experimental parameters were chosen for the best compromise between minimizing the background, optimizing the resolution, and maximizing the scattered intensity: $H \simeq 5\mathrm{kOe}$ ($Q_0 \simeq 0.011 ~ \text{\AA}^{-1}$), $\lambda _{\mathrm{N}} \simeq 7~\text{\AA}$ ($\pm 10\%$). Based on the wavelength spread and beam collimation parameters ($12.8~\mathrm{m}$ collimation, $30~\mathrm{mm}$ source aperture, $5~\mathrm{mm}$ sample pinhole), the angular resolution along the rocking angle was estimated to be $\Delta \theta_\mathrm{res} \sim 0.086^\circ$ (FWHM). The applied magnetic field was parallel to crystal \textbf{c} axis and nearly parallel to the beam. A special sample holder, equipped with a set of coils, allowed measuring the AC susceptibility response \textit{in situ}.

\section{Results and discussion}
\label{sec:results}

\subsection{History effects in the effective pinning}

\begin{figure}[t]
    \centering
    \includegraphics[width=0.99\linewidth]{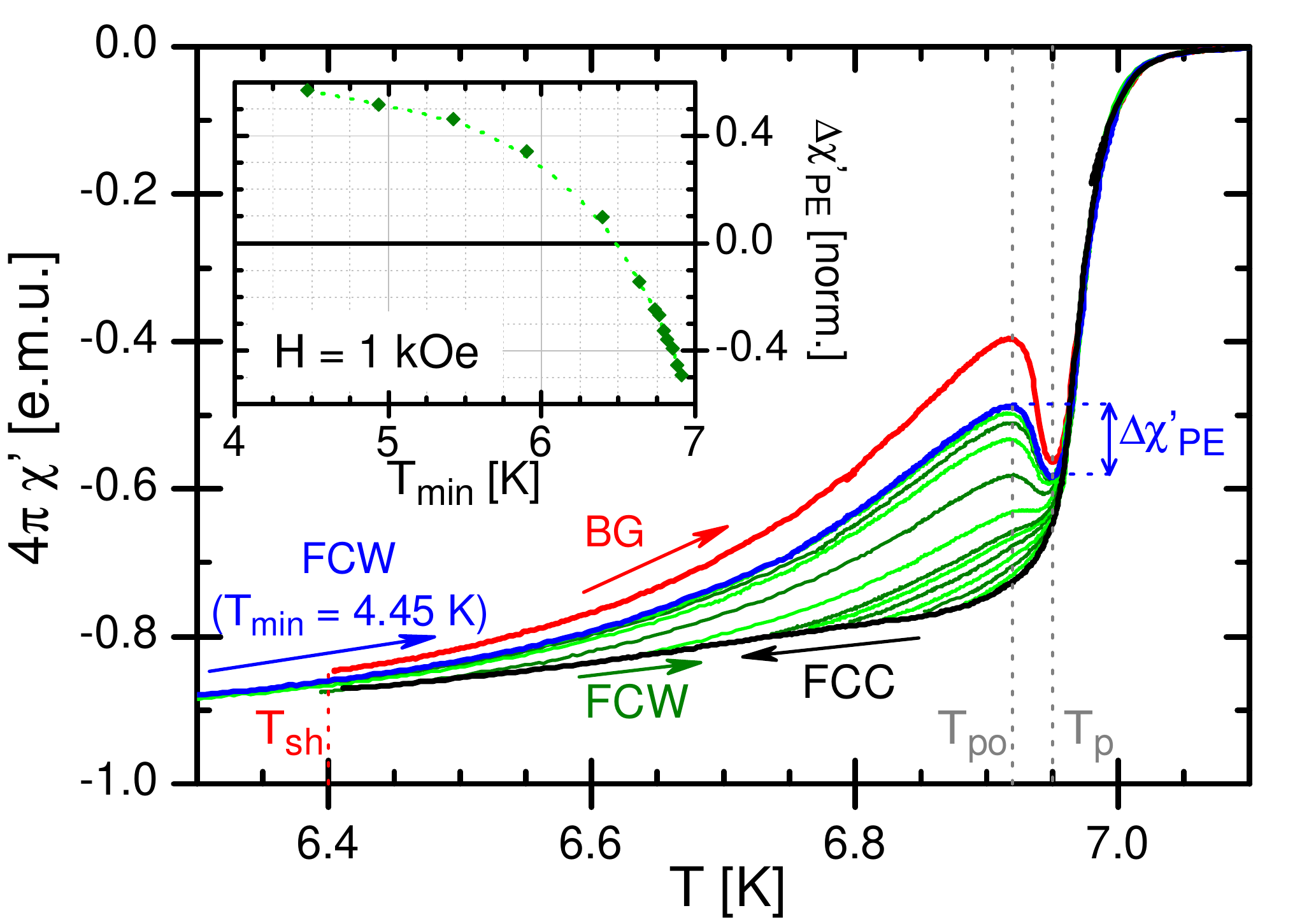}
    \caption{In-phase linear AC susceptibility $\chi'(T)$ recorded during a FCC process (black curve), after shaking the system at $ T_\mathrm{sh} < T_\mathrm{po}$ (red curve), and during various FCW from different minimum temperatures $T_{\min }$ (green and blue curves). Arrows indicate the direction of temperature variation; vertical dotted lines indicate relevant temperatures (see text). FCW processes display lower effective pinning than FCC without any dynamic assistance.  Inset: Normalized height of the PE (Eq. \ref{eq:peakheight}, see text) measured in FCW processes as a function of the lowest temperature reached, $T_{\min}$. The dotted line is a guide to the eye.}
   \label{fig:ac_susc1}
\end{figure}

Linear AC susceptibility curves measured after different thermal and dynamic histories are shown in Figure \ref{fig:ac_susc1}. As reported in previous works, in a field-cooled cooling process (FCC, black curve), the VL remains trapped in a strongly pinned configuration and $\chi'(T)$ displays no PE. After shaking the VL at $T_\mathrm{sh}$ well below the onset of the PE ($T_{\mathrm{po}}$), an unambiguous rise in $\chi'$ (red curve) signals an increment in $\lambda _{\mathrm{AC}}$ and thus a softening of the effective pinning. In addition, a strong PE is observed upon warming the shaken VL. Similar dynamic assistance can also be provided by a transport current or a sudden increase/decrease of the magnetic field (e.g., zero-field-cooled configurations). Recent SANS \cite{Marziali2015} and STS \cite{Zehet2015,Ganguli2015} observations, in agreement with pioneer experiments,\cite{Yaron1994} corroborated that the decrease of the effective pinning after applying a shaking field, or an equivalent dynamical assistance, is correlated with a bulk ordering of the VL where metastable supercooled VL dislocations are removed.

Certain thermal protocols can also reduce effective pinning without any dynamic assistance.\cite{Pasquini2008}  After cooling the system down to sufficiently low temperatures, $\chi'(T)$ curves measured during subsequent warming processes (FCW, green curves in Figure \ref{fig:ac_susc1}) display the PE and a clear reduction of effective pinning with respect to the original FCC process. To what extent pinning is softened depends on the lowest temperature reached $T_{\mathrm{min}}$. For a clearer comparison among the various thermal histories, we define the height of the PE as
\begin{equation}
\Delta \chi'_\mathrm{PE} = \chi'(T_\mathrm{po}) - \chi'(T_\mathrm{p}),
\label{eq:peakheight}
\end{equation}
where $T_\mathrm{po}$ is the local maximum of $\chi '(T)$ (matching the onset of the PE) and $T_\mathrm{p}$, the local minimum (maximum effective pinning) when the PE is observed. In the inset of Figure \ref{fig:ac_susc1}, $\Delta \chi'_\mathrm{PE}$, normalized by the corresponding height for a VL shaken in the Bragg Glass phase, is shown as a function of $T_{\mathrm{min}}$. The PE occurs (i.e., $\Delta \chi'_\mathrm{PE} > 0$) only after the sample was cooled below a threshold temperature ($\sim 6.5~\mathrm{K}$ in the example shown in Figure \ref{fig:ac_susc1}). Once the PE develops, $\Delta \chi'_{PE}$ increases with decreasing $T_{\mathrm{min}}$. The curves are independent of cooling and warming rates and whether linear susceptibility was measured during the cooling process or not. FCW curves warmed from the lowest accessible temperatures in our experiments display marginally different heights and, for all $T_\mathrm{min}$, $\Delta \chi'_{PE}(T_\mathrm{min}) < \Delta \chi'_{PE}(\mathrm{BG})$. Hence, the effective pinning reduction due to thermally cycling the system approaches a limit as $T_\mathrm{min} \rightarrow 0$, without displaying as strong a reduction as that observed after dynamically assisting the system.

\subsection{History effects in the VL structure}

\begin{figure}[b]
  \centering
    \includegraphics[width=0.99\linewidth]{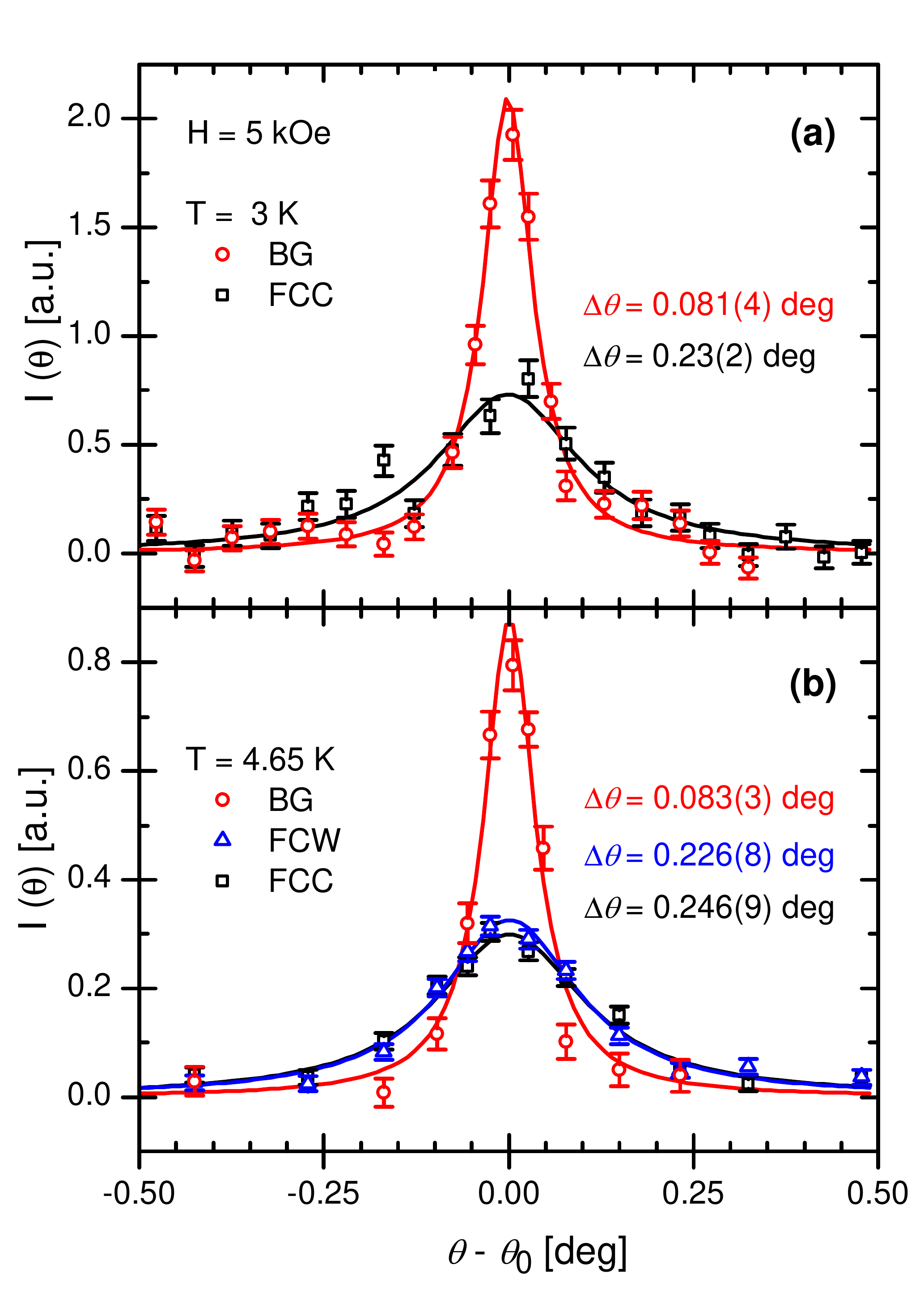}
    \caption{Rocking curves from SANS measurements after different thermal and dynamic histories recorded at the base temperature $3~\mathrm{K}$ (a) and at $4.65~\mathrm{K}$ (b): FCC (black open squares), VL shaken at $5.6~\mathrm{K}$ (red open circles) and FCW from $3~\mathrm{K}$ (blue open triangles). Continuous lines are Lorentzian fits with full width at half maximum $\Delta \theta$.
    }
    \label{fig:RCs}
\end{figure}

Thermal-history-dependent effective pinning has been ascribed to \textit{spontaneous VL ordering} promoted by the faster strengthening of vortex-vortex interactions with respect to pinning forces as the system is cooled in the BG phase.\cite{Pasquini2008, Li2006} Following the observed connection between the effective pinning reduction and the removal of dislocations after dynamical assistance, spontaneous annihilation of dislocations might also be expected to occur during the proposed thermal ordering. A structural characterization through SANS experiments was performed in order to assess the role of dislocations in the observed hysteretic pinning.

If dislocations were removed as a result of thermal cycling, larger correlation lengths $\zeta$ due to a reduction of the VL dislocation density should be observed as narrower rocking curves (RCs). In Figure \ref{fig:RCs}, we show RCs recorded at the base temperature ($3 ~ \mathrm{K}$, panel a) and at a higher temperature ($4.65 ~ \mathrm{K}$, panel b). RCs acquired after shaking the VL at $5.6~\mathrm{K}$, in the BG phase (red open circles), display resolution-limited Bragg Peaks at both temperatures, indicating correlation lengths much larger than our resolution. Their corresponding full width at half maximum (FWHM) was estimated from Lorentzian fits to the data, obtaining $\Delta \theta = 0.081(4)^\circ$ and $0.083(3)^\circ$, respectively. These values are slightly smaller and imply a better resolution limit ($\zeta_z \lesssim 40~\upmu\mathrm{m}$) than the rough estimate $\Delta \theta_\mathrm{res} \simeq 0.087^\circ$ (Sec. \ref{sec:details}). Conversely, the FCC RCs (black open squares) are clearly wider than $\Delta \theta_\mathrm{res}$, indicating a VL with a high density of dislocations. Here, positional order is lost beyond a characteristic length $\zeta_z = 4.9(2)~\upmu\mathrm{m}$.\cite{Marziali2015} A FCW RC, recorded after warming a FCC configuration from the base temperature to $4.65 ~ \mathrm{K}$, is also shown in Figure \ref{fig:RCs}b  (blue open triangles). We found this FCW RC to be near identical to the FCC RC, with $\zeta_z = 4.5(2) ~ \upmu\mathrm{m}$. This result indicates no substantial modification of $\zeta_z$ during the cooling-warming process up to this temperature.

\begin{figure}[b]
\centering
\includegraphics[width=0.99\linewidth]{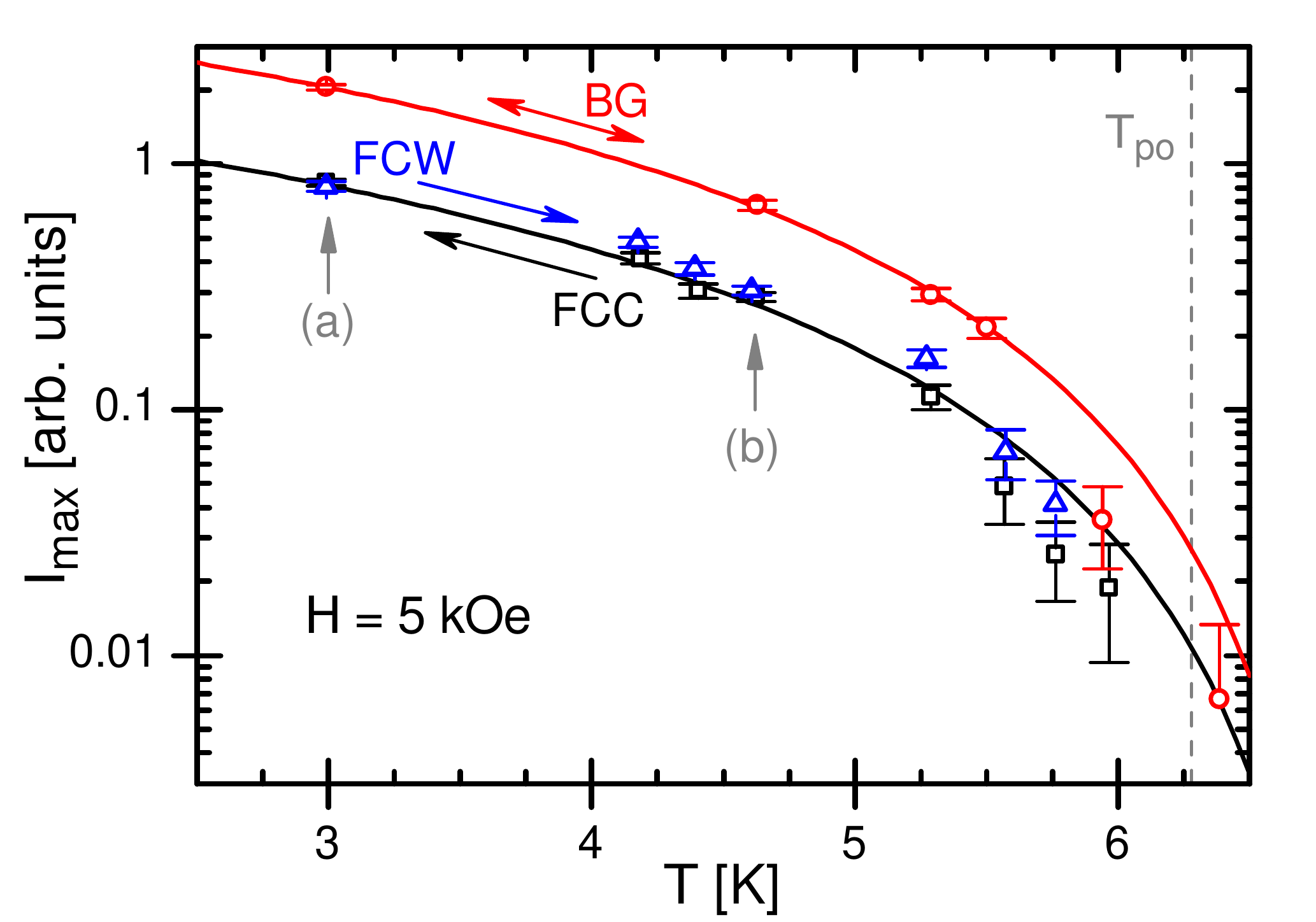}
\caption{Maximum BP intensity, $I_{\max}(T)$, recorded during cooling/warming processes after different dynamic histories: VL shaken in the BG phase (red dots), FCC (black squares), and FCW (blue triangles). Continuous lines show the expected $F^{2}(T)$-like dependence from Eq. \ref{London model}, using an effective vortex core size (see text). Arrows indicate the direction of temperature evolution in the three processes (labelled FCC, FCW, and BG) and the intensities corresponding to RCs shown in Figure 2 (labelled a and b). A vertical dashed line marks $T_\mathrm{po}$.
}
\label{fig:sans_results}
\end{figure}

As explained in Sec. \ref{sec:sansvortex}, measuring full RCs at higher temperatures was not feasible. Instead, $I_\mathrm{max}(T)$ was acquired at several temperatures for the various VL configurations (Figure \ref{fig:sans_results}). The fact that FCC and BG RCs recorded at each temperature have the same integrated intensity within the typical precision of the SANS measurements ($10\%$) strongly supports using $I_\mathrm{max}$ as probe for changes in $\zeta_z$. As long as $\zeta_z$ is beyond the experimental resolution, the maximum BP intensity scales as the integrated intensity. Therefore, for the ordered Bragg Glass BPs (red dots), $I_\mathrm{max}(T)$ is expected to decay as the square of the single-vortex form factor $F^{2}(T)$, at least for low temperatures. $\mathrm{NbSe}_{2}$ is a multi-band superconductor\cite{multiband} with six-fold star shaped vortex core;\cite{Suderow} therefore, models predicting $F^{2}(T)$ based on the London theory should be used with caution. Still, using $c = 0.12$ and estimating the vortex core size as
\begin{equation}
    \xi(T) = \sqrt{ \frac{\phi_{0}}{2\pi H_{c2}(T)} },
    \label{eq:xi_Hc2}
\end{equation}
the modified London model (Eq. \ref{London model} and red line in Figure \ref{fig:sans_results}) is in good agreement with the experimental $I_\mathrm{max}(T)$ of the resolution-limited BPs. Here, $H_{c2}(T)$ was estimated from the onset of the diamagnetic response,\cite{SIMarziali2015} and $\lambda_\mathrm{L}(T)$ was extracted from direct experimental data.\cite{Fletcher2007} The small value of $c$ may be a consequence of overestimating $\xi$ in Eq. \ref{eq:xi_Hc2} or due to a lesser effect of the finite core size in this system. The deviation of experimental data from the expected decay around $T \sim 6~\mathrm{K}$ (still in the ordered phase) may be due to a decrease of the elastic BG correlation length below experimental resolution.\cite{Klein2001}

The same model and parameters were used for the FCC data, with the expected $F^{2}(T)$  scaled to match $I_\mathrm{max}$ at the lowest temperature (black line in Figure \ref{fig:sans_results}). As with the BG, there is no apparent deviation from the expected decay up to $T \sim 6 ~ \mathrm{K}$, suggesting no substantial modification of $\zeta_z$ nor of the mean dislocation distance\cite{Marziali2015} up to this temperature. This result is consistent with the fact that FCC RCs recorded at both temperatures displayed similar $\Delta \theta$, and extends to higher temperatures. Moreover, in spite of the large decrease in the effective pinning during warming processes (Figure \ref{fig:ac_susc1}), $I_\mathrm{max}(T)$ corresponding to FCC and FCW configurations show only minor differences up to the highest temperature at which we were able to acquire neutron data. Therefore, correlation lengths of a bulk dislocated metastable FCC VL are not substantially modified after a cooling/warming process in the BG phase. This behavior is qualitatively different to that observed in Nb single crystals, where a gradual decrease of $\zeta_z$ during FC processes in the BG phase has been reported.\cite{Daniilidis2007} Our results suggest that the spontaneous \textit{thermal ordering} does not involve significant annihilation of VL dislocations, in contrast with the strong change in the dislocation density obtained after an assisted \textit{dynamic ordering}.

\subsection{Towards a consistent description}

Hysteretic responses in glasses are ascribed to energy barriers that may trap the system in metastable configurations. Our observations suggest that there are mainly two different types of energy barriers: 1) those that are easily overcome and vanish as the relative weight between competing interactions changes with temperature, and 2) those exceeding the available thermal energy and therefore requiring dynamical assistance to annihilate topological defects and drive the VL to the ordered Bragg glass phase at low temperature. These two kinds of energy barriers explain our results and reveal an interesting underlying physics. In a metastable VL, the size of the correlated volumes, determined by the dislocation density, are not drastically affected by the thermal history. However, at low temperature, elastic vortex-vortex interactions prevail over vortex-pinning interactions, producing short-range vortex accommodation inside each correlation volume. In this local reorganization, the VL overcomes lower energy barriers, leading to spontaneous local ordering. This subtle accommodation can drastically affect the VL effective pinning and, thus, linear AC response, but would not significantly affect the neutron scattering pattern.

\begin{figure}[tb]
   \centering
    \includegraphics[width=0.99\linewidth]{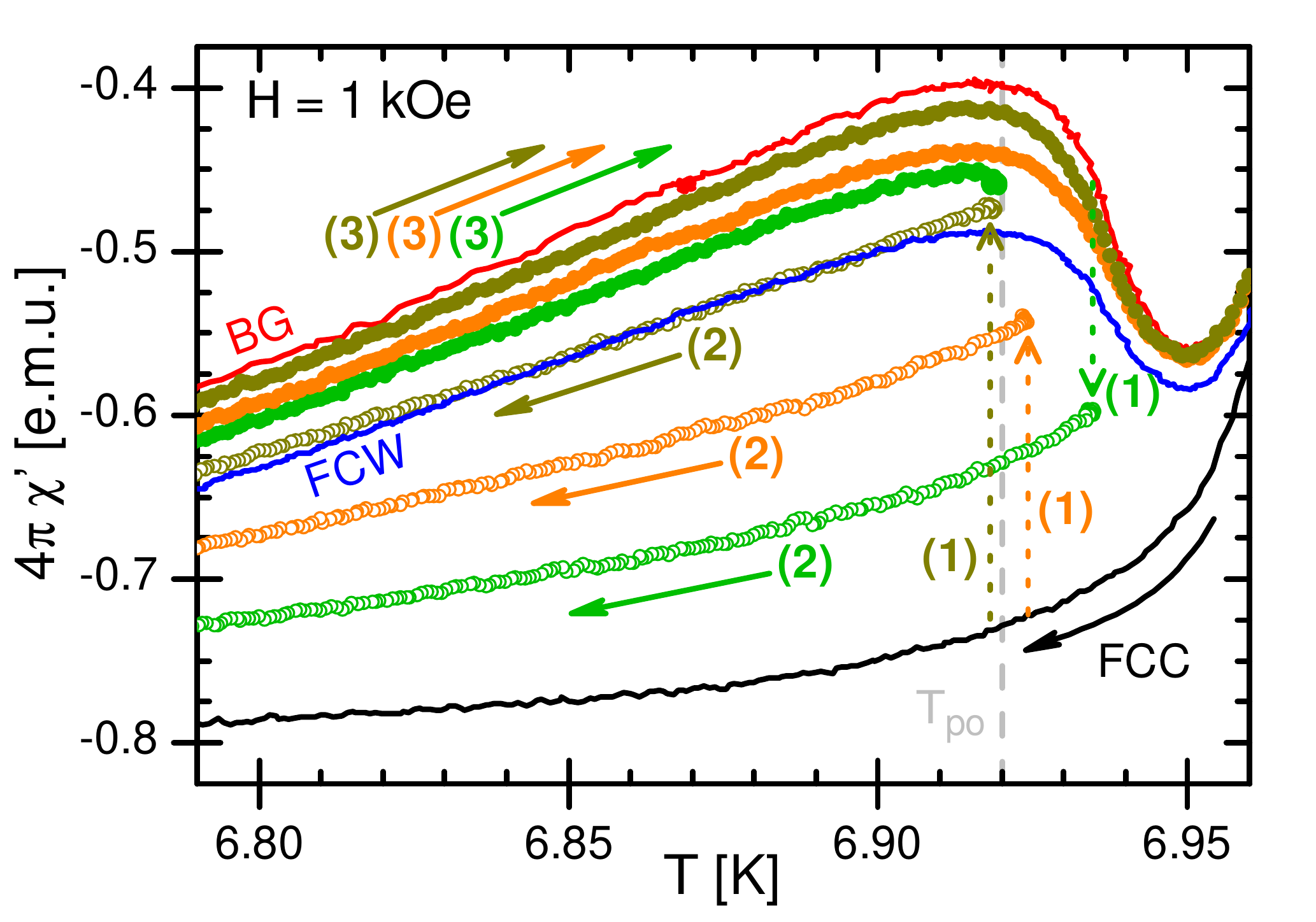}
    \caption{AC response measured during cooling (open symbols) and immediately warming (full symbols) VL configurations with intermediate disorder. Each of these configurations (dark yellow, orange and green symbols) was prepared by shaking the VL at a different temperature in the transitional region, near $T_\mathrm{po}$ (represented by arrows in dotted lines). Arrows in continuous lines indicate the direction of temperature evolution and numbers in parenthesis, temporal order. Continuous lines reproduce FCC (black), FCW (blue) and BG (red, shaken well below $T_\mathrm{po}$) curves for reference. Each warming response is still intermediate between FCW and BG responses.}
    \label{fig:ac_susc2}
\end{figure}

Results presented in Secs. \ref{sec:results}A and \ref{sec:results}B support the proposed scenario based on observations in extreme situations: the most ordered (BG) and the most disordered (FC) VL configurations. To further test the consistency of our picture, we analyzed the thermal cycling response of VL configurations initially stabilized with various \textit{intermediate} dislocation densities. These configurations were prepared by shaking the system at various temperatures near $T_\mathrm{po}$, within the transitional region.\cite{Marziali2015} The dislocation density is expected to increase as the shaking temperature approaches $T_\mathrm{p}$.\cite{Pasquini2008} Figure 4 shows the AC responses corresponding to cooling-warming processes for various VL configurations obtained (green, orange, and dark yellow symbols). We compare these results with the thermally cycled response of an initially ordered VL without dislocations, shaken well below $T_\mathrm{po}$ (red curve), and a more disordered FCW VL, not shaken (blue curve). Each intermediate cooling response depends on the shaking temperature.\cite{Pasquini2008} After cooling the system to the same $T_{\mathrm{min}}$, all warming curves reveal an important reduction of the effective pinning with respect to the corresponding cooling process, as in FCC-FCW cycles. Moreover, each warming response is intermediate between the FCW and the BG response, and correlated with the shaking temperature that determined each dislocation density. Therefore, these results indicate that there remains some memory of the original dislocation density after thermally cycling the system. This dislocation density memory is in agreement with recent direct STS observations of supercooled configurations,\cite{Ganguli2016} in which dislocations were still present upon cooling the system to low temperatures. The fact that pinning is reduced while dislocations remain supports our proposing two kinds of energy barriers.

It is particularly interesting whether both kinds of barriers play an important role in the PE. The spontaneous VL disordering responsible for the increase in the effective pinning in the transitional region (sudden drop in $\chi'(T)$ above $T_{\mathrm{po}}$) has been a subject of controversy for decades. The first simple explanation, given by Pippard,\cite{Pippard1969} was based on the accommodation of vortices to the random pinning center distribution: As $H_{c2}$ is approached, the rigidity of the flux-line lattice drops to zero faster than the pinning strength. This allows the flux lattice to adapt more easily to the underlying quenched disorder, so that the effective pinning is increased. Subsequent models by Larkin and Ovchinnikov\cite{LO} provided quantitative relations between the elastic VL modulus, pinning forces, and the effective critical current density. These models lost attention years later, when the role of VL dislocations in increasing the effective pinning at the ODT became evident. This new observation suggested that, near the depinning transition, plastic creep and plastic dynamics largely boost the effective critical current density and reduce VL mobility.\cite{Kierfel2000} Still, our results show that the dislocation density only partially determines the effective pinning. May both descriptions of the PE hold under different conditions?

\begin{figure}[tb]
    \centering
   \includegraphics[width=0.99\linewidth]{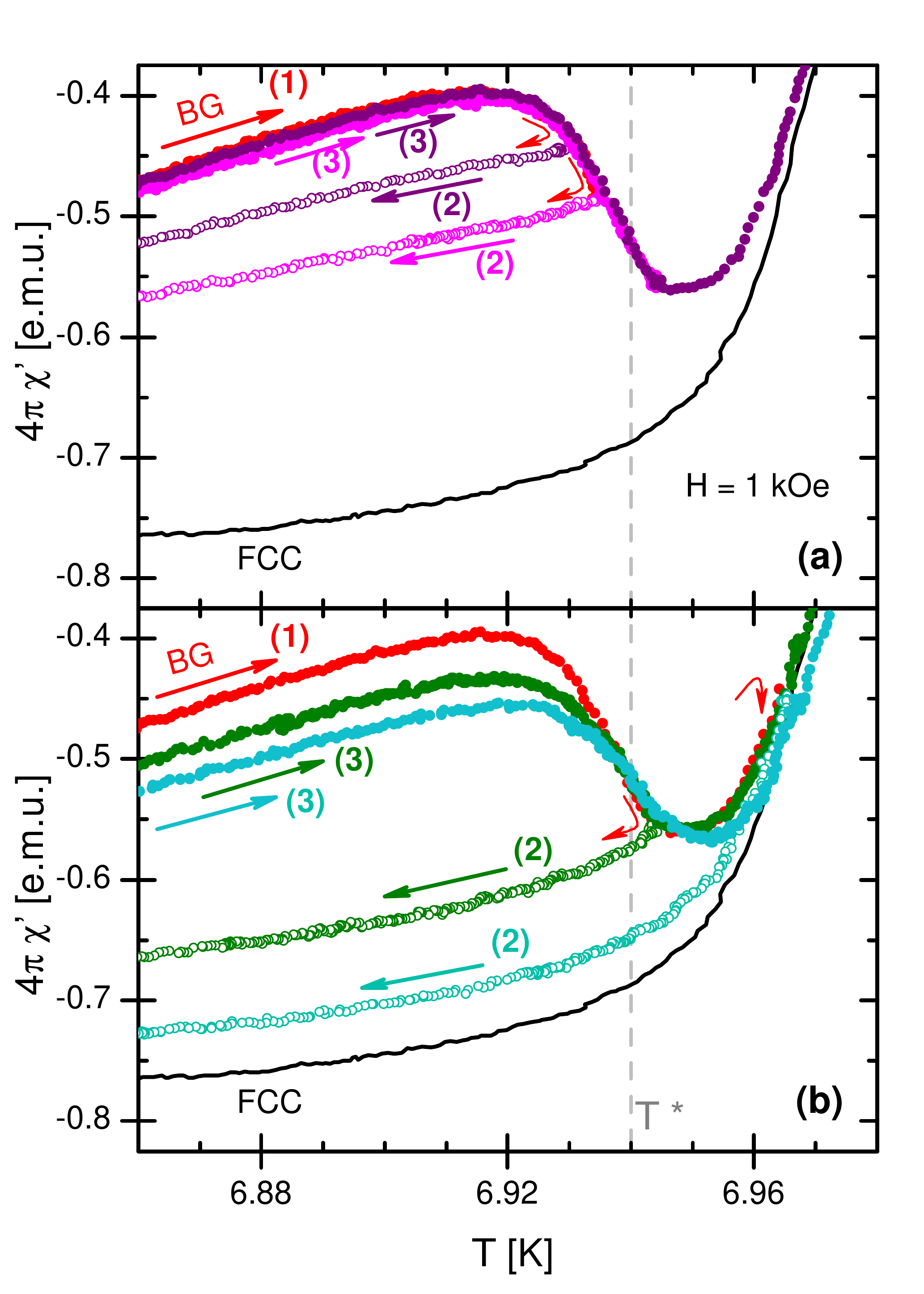}
   \caption{AC response measured while thermally cycling an initially ordered VL. Straight arrows indicate the direction of temperature evolution and red curved arrows mark turning points. In each sequence, (1) a \textit{fresh} BG was warmed to $T_{\max}$ (red symbols), (2) cooled to $4.5~\mathrm{K}$ (open symbols), and (3) warmed again (full symbols). There is a temperature $T^{*}$ (vertical dashed line) such that, if  $T_{\max }<T^{*}$, the BG response is recovered (panel a), but \textit{memory} is lost when $T_{\max }>T^{*}$ (panel b).}
   \label{fig:ac_susc3}
\end{figure}

In order to clarify the spontaneous thermal disordering above $T_{\mathrm{po}}$, we conducted a systematic study of history effects in the AC response. The linear AC response recorded in various thermal cycles, each starting from an ordered VL, is shown in Figure \ref{fig:ac_susc3}. After interrupting the warming process (full red dots) in the transitional region, all cooling curves show an intermediate degree of pinning, correlated with the maximum temperature reached. In Figure \ref{fig:ac_susc3}a, cooling processes from two different upper temperatures (open pink and purple dots) are shown as examples. Strikingly, the subsequent warming processes (full pink and purple dots) recover the reversible BG response without any dynamic assistance. Identical responses suggest that the system evolves in such a way that the \textit{spontaneous disordering} produced in the PE region is recovered by a \textit{spontaneous ordering} during the cooling process. A similar memory effect has been reported in previous SANS experiments.\cite{Marziali2015} There we showed that a low-temperature RC measured after thermally cycling an ordered VL through the transitional region remained beyond the resolution limit. This implies that solely warming the system up to the transitional region does not create a significant number of dislocations, if any at all. In addition, superheated configurations free of dislocations above $T_{\mathrm{po}}$ have been recently observed in STS experiments.\cite{Ganguli2016} Altogether, these experiments evidence that the PE cannot be a consequence of the proliferation of dislocations exclusively. In particular, the effective pinning increase displayed \textit{right above} $T_{\mathrm{po}}$ would only involve the elastic accommodation of vortices to the random quenched disorder, which can be easily undone by cooling the system. 

Thermal history effects associated to the creation of dislocations are likely to become apparent at higher temperatures. Consequently, the memory of the dislocation density will be lost if the system is warmed beyond some characteristic temperature $T^*$. The number of spontaneous dislocations (and disclinations) is expected to increase when approaching the disordered phase,\cite{Ganguli2015} and a completely amorphous VL would lie above the spinodal line.\cite{Xiao2004} Panel \ref{fig:ac_susc3}b shows the evolution of the AC response starting from an ordered configuration obtained by shaking the VL in the BG phase (red) and thermally cycling the sample reaching temperatures well above $T_{\mathrm{po}}$: near the maximum effective pinning temperature $T_{\mathrm{p}}$ (open green symbols) and above $T_{\mathrm{p}}$ (open cyan symbols). It can be seen that the BG response is no longer recovered during the second warming processes (full green and cyan symbols), and the memory is lost. Within the context of the proposed picture, our results support a novel scenario to understand the PE: A dynamic assistance would generate VL dislocations immediately above $T_{\mathrm{po}}$, breaking metastable ordered configurations which would otherwise remain up to $T^*$. In contrast, above $T^*$, plastic energy barriers are lower and thermal energy alone is sufficient to overcome them. 

\section{Conclusions}

\label{sec:conclusions}

We have studied thermal and dynamic history effects in the VL in clean $\mathrm{NbSe}_2$, inspecting the evolution of effective pinning by means of AC susceptibility techniques and observing the VL bulk structure in SANS experiments. The reported results suggest a novel scenario to describe the order-disorder transition and the Peak Effect anomaly. Additionally, the proposed framework helps to clarify the controversial relationship between the VL structure and effective pinning. In the proposed picture, the interplay between two different types of energy barriers is the origin of the history and memory effects in the effective pinning potential. On one hand, high plastic barriers impeding the annihilation or creation of topological defects, that would otherwise reduce the total VL energy, require dynamic assistance to be overcome. On the other hand, low elastic barriers prevent the VL from accommodating the pinning landscape at the local level. These barriers are easily overcome and vanish as the relative weight between competing interactions changes with temperature. This subtle elastic arrangement would favor the elastic vortex-vortex interaction at low temperatures and the vortex-pinning interaction in the transitional region. This constitutes the main ingredient for the spontaneous PE and the observed thermal hysteresis in the linear AC response.

In our scenario, thermal history does not drastically affect the size of the correlated volumes, which is mainly determined by the mean dislocation density, but clearly affects vortex arrangements within the correlated volumes. This provides an explanation for different pinning intensities in configurations showing similar low-temperature SANS rocking curves. Quenched dislocations in field cooling protocols are responsible for the wide RCs recorded in SANS experiments\cite{Marziali2015} and supercooled disordered VLs observed by STS at low temperature.\cite{Ganguli2016}  High energy barriers also prevent the creation of dislocations in metastable superheated ordered configurations in the transitional region (between the PE onset temperature $T_\mathrm{po}$ and a characteristic temperature $T^* \lesssim T_\mathrm{p}$), preserving narrow RCs and memory in the effective pinning after thermally cycling the system below $T^*$. Beyond $T^*$, plastic energy barriers vanish, favoring the spontaneous creation of topological defects. In STS experiments performed in doped $\mathrm{NbSe}_2$ crystals, a topological transition has been identified at $T \sim T_\mathrm{p}$, above which stable VL configurations are highly disordered with proliferation of VL dislocations and disclinations.\cite{Ganguli2016} The vanishing of plastic barriers above $T_\mathrm{p}$ is possibly associated with this topological transition. Although the proposed picture is consistent with the experimental evidence and physically sound, a formal theoretical treatment is needed. For instance, hysteretic Campbell response in the presence of strong pinning centers has been recently demonstrated through a simple model.\cite{Willa2015} A similar approach for VLs in weak pinning landscapes should help to better understand the connection between thermal history leading to metastable VL configurations and hysteretic Campbell responses.

An important remark: When vortex displacements involve larger distances, dislocations are created, annihilated, and/or redistributed, erasing metastable configurations and its related memory effects. Hence, the proposed picture and its underlying physics cannot be accessed through invasive techniques. This explains the disagreement between observations by linear AC susceptibility or ultrafast transport techniques and conventional transport experiments, in which VL self-reordering and edge contamination are also present.\cite{Paltiel2000}

Finally, we mention that partial spontaneous removal of dislocations in the cooling process of a disordered VL cannot be ruled out. The FC VL at low $T$ displays bulk orientational order and the mean dislocation distance, estimated from the width of the RCs, is much larger than $a_0$. Therefore, it is expected that successive thermal cycles in an adequate range of temperatures could eventually help the progressive removal of dislocations without dynamic assistance. Direct observations in the full temperature range and a systematic study of different thermal cycles may enable the corroboration of the proposed picture.

\section*{Acknowledgements}
MMB, VB, and GP were supported by CONICET under Grant PIP 536 and Universidad de Buenos Aires; ERL and MRE were supported by the U.S. Department of Energy, Office of Basic Energy Sciences, under Award No. DE-FG02-10ER46783. We are grateful to G. Cuello and G. Garbarino for their collaboration in submitting the experimental proposal and their assistance with logistics, and to G. Nieva for providing the $\mathrm{NbSe}_2$ samples.

\end{document}